# Deterministic Terahertz wave control in scattering media

Vivek Kumar[1], Vittorio Cecconi[1], Luke Peters[1], Jacopo Bertolotti[2], Alessia Pasquazi[1], Juan Sebastian Totero Gongora[1], Marco Peccianti[1]*

[1] Emergent Photonics Lab (EP*ic*), Department of Physics and Astronomy, University of Sussex, Brighton, BN1 9QH, U.K.

[2] Department of Physics and Astronomy, University of Exeter, Exeter, Devon EX4 4QL, UK

**ABSTRACT:** Scattering-assisted synthesis of broadband optical pulses is recognized to have a cross-disciplinary fundamental and application importance. Achieving full-waveform synthesis generally requires means for assessing the instantaneous electric field, i.e. the absolute electromagnetic phase. These are generally not accessible to established methodologies for scattering-assisted pulse envelope and phase shaping. The lack of field sensitivity also results in complex indirect approaches to evaluate the scattering space-time properties. The terahertz frequency domain potentially offers some distinctive new possibilities thanks to the availability of methods to perform absolute measurements of the scattered electric field, as opposed to optical intensity-based diagnostics. An interesting conceptual question is whether this additional degree of freedom can lead to different types of methodologies towards wave shaping and to a direct field-waveform control. In this work, we theoretically investigate a deterministic scheme to achieve broadband, spatiotemporal waveform control of terahertz fields mediated by a scattering medium. The direct field access via Time-Domain Spectroscopy enables a process in which the field and scattering matrix of the medium are assessed with minimal experimental efforts. Then, the illumination conditions for an arbitrary targeted output field waveform are deterministically determined. In addition, the complete field knowledge enables reconstructing field distributions with complex phase profiles, as in the case of phase-only masks and optical vortices, a significantly challenging task for traditional implementations at optical frequencies based on intensity measurements aided with interferometric techniques.

**Graphical Abstract**

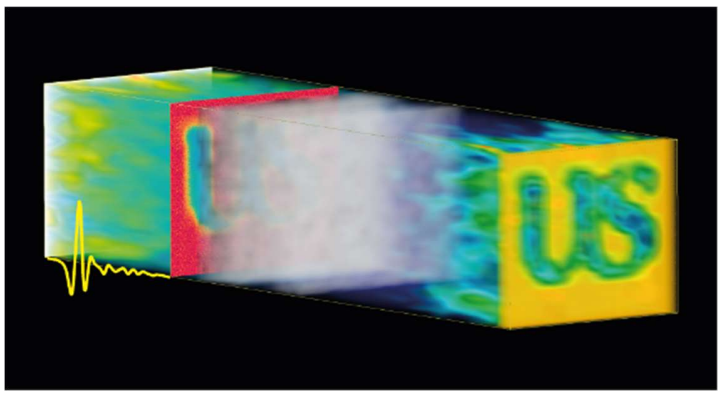

## Introduction

The propagation of waves in a scattering medium results in complex space-time interference patterns, i.e., in a complex time and position-dependent response at the output [1,2]. These phenomena are ubiquitous features in the physics of random wave propagation and significantly impact applications in several domains ranging from electromagnetic to acoustic, mechanical, and matter waves[2,3]. For instance, in optical imaging, the random propagation of light rapidly reduces the image fidelity in deep biological tissue characterization[4]. As such, the performance of state-of-the-art microscopes is traditionally affected by the ineliminable dynamical turbidity in the samples.[5,6]

Although light scattering is usually considered an impediment, it is not necessarily accompanied by an irreversible loss of information[7]. By leveraging this principle, researchers have recently developed a broad range of wavefront-shaping techniques to control complex light propagation through a scattering medium[8,9]. The basic principle is to spatially modulate the wave impinging onto the medium to harness the scattering-induced amplitude and phase distortions. Recently, various approaches towards optical wavefront compensation based on feedback[10], guided-stars[11], and memory-effect[12] have been demonstrated in different disciplines. Those methodologies have enabled the manipulation of scattered waves for the re-focusing and imaging applications. Although approaches based on the iterative optimization of the scattered field rely on technically simple implementations, they fundamentally operate without direct knowledge of the scattering medium. As such, a specific optimization process provides little clues for a different one, and convergence is usually established solely by the inability to reduce an error function further. Deterministic approaches overcome this limitation. They rely on the knowledge provided by measuring the optical transfer matrix of the medium [13,14]. Deterministic methods first measure the scattered light field corresponding to different sets of amplitude[15,16] or phase[17,18] illumination patterns (preferably forming an orthogonal set). The measurements are then combined to achieve a single-step illumination retrieval for a desired optimized wavefront through numerical inversion.

## 1. The challenge of field-wave synthesis using random media

Within the process of exploiting a random media for space-time wave synthesis, one can argue that the knowledge of the transmission matrix is insufficient. While the transmission matrix via complex spectral interferometry approaches, synthesis requires the prior knowledge of the source. If the absolute phase profile of the source pulse is not known, it is also unknown in the scattered field.

Interestingly, when the detection can resolve the instantaneous field dynamics for a large set of spatially modulated fields, one can trivially target a new scattered waveform as a simple combination of the field scattered by different illuminations without directly referring to the scattering matrix. Indeed, powerful wave-synthesis approaches in optics do not generally rely upon absolute phase knowledge [18]. This rationale, for example, is one of the critical accelerating factors for optical frequency combs technology[19]. Conversely, persuasive, popular nonlinear pulse diagnostics[20–22] (e.g. FROG or SPIDER) do not provide access to the instantaneous field.

## 2. A full time-domain field-approach

The ability to perform a complete time-domain detection brings a conceptual difference: by introducing a sparse-light modulation (as in the practice of random media functionalization) for each spatially orthogonal illumination $p_i$ one can detect the corresponding space-time waveform $E_i^+(x_o, t)$. These independently transmitted waveforms can, in principle, be used to decompose any desired space-time waveform $E_T^+(x_o, t)$ at the output as a linear superposition:

$$E_T^+(x_o, t) = \sum_i c_i E_i^+(x_o, t) \tag{1}$$

where $c_i$ are complex-valued expansion coefficients. Once one determines the set of values of $c_i$ experimentally achievable, this can provide access to the spectrum of available waveforms. This approach does not require specific access to the source waveform, which is however trivially accessible via a time-sensitive detection and would grant access to the scattering matrix.

In this context, terahertz time-domain spectroscopy (THz-TDS) is a mature and established technique capable of fully resolving the electric field oscillations in a broadband pulse[23], i.e. providing full complex spectral field knowledge[24]. The scientific question is whether a THz-TDS can be exploited to develop a deterministic approach to waveform synthesis (closely related to ultrasound or radiofrequency approaches[25,26]). The idea is to extract sufficient information to obtain access to any scattering-allowed output field. The large relative THz bandwidth available in TDS embodiments (normally exceeding a decade) allows easily spanning a wide range of single and multiple scattering regimes for a given sample given scattering element[27,28]. On the practical side, the relatively large wavelength of THz waves (spanning from roughly 30 $\mu m$ to 3 mm) suggests that the typical subwavelength scales of scattering phenomena are significantly more accessible in experimental platforms, when compared to optical embodiments.[29–31]

A general downside in implementing THz wavefront control methods is, however, the limited availability of wavefront shaping devices.[32,33] Besides, the use of diffraction-limited systems at long wavelengths (which fixes the pattern resolution[34,35]) is undesirable because the experimental setting usually does not involve samples several orders of magnitude larger than the wavelength, trivial condition in optics. This results in a relatively small number of modes that can be independently excited in a scattering structure with far-field illumination.[36] Very recently, the nonlinear conversion of structured optical beams has emerged as a promising approach towards spatial light modulation of THz waves.[37,38] The combination of nonlinear THz pattern generation and time-resolved field detection, in particular, has enabled the development of hyperspectral THz imaging with deeply subwavelength imaging resolution.[39] In essence, placing an object in the near-field of a nonlinear optical-to-THz converter makes it possible to produce terahertz illumination patterns with fine spatial features approaching the optical (i.e. the pump) diffraction limit.

In this work, we explore this framework in connection with scattering-assisted waveform synthesis, introducing the field equivalent of the traditional spatiotemporal focusing and image retrieval. We explore scenarios extremely challenging in optics, which include retrieval of field distributions with complex phase profiles, such as phase-only masks and optical vortices.

In our approach, we expand the complex-valued, coherent transfer matrix of the scattering medium using an orthogonal Walsh-Hadamard decomposition of the near-field THz illumination[40]. We leverage this knowledge to perform a direct single-step inversion using a constraint least-square optimization approach compatible with realistic experimental conditions.[41]

## 3. Methods: model definition and simulation setup

We define the input/output field relation in term an impulse response $T_x(x_o, x', t, t')$[42], as:

$$E^+(x_o, t) = \int \int T_x(x_o, x', t, t') E^-(x', t') dx' dt' \quad (2)$$

where $E^-$ and $E^+$ denote the spatio-temporal electric field distribution just before and after the scattering medium. To lighter the notation, we define $x_o$ and $x'$ as the one-dimensional representation of the input and output planes, respectively. In the frequency domain $\omega$, Eq. (1) reads:

$$E^+(x_o, \omega) = \int \widetilde{T_x}(x_o, x', \omega) E^-(x', \omega) dx' \quad (3)$$

where $E^+(x_o, \omega)$, $E^-(x', \omega)$ are the time-Fourier transforms of the input and output fields.

Following standard approaches, we rewrite the continuous relationship set Equations (2-3) in a discrete scalar transfer matrix formalism, where the response of the scattering medium for each incident frequency is described by an $M \times N$ field-based, random transmission matrix $T_{mn} \in C^{M \times N}$,[43]. We divide the output and input planes into $M$ and $N$ spatial independent segments (corresponding, e.g., to the physical pixels on the input wavefront-shaping and output imaging devices) and $K$ spectral modes, a representation that is well suited for experiments. For a given $k$-th frequency $\omega_k$ the relationship between the THz fields $E_m^+(\omega_k)$ and $E_n^-(\omega_k)$ at the $n$-th input and $m$-th output pixels reads as follows:

$$E_m^+(\omega_k) = \sum_n T_{mn}(\omega_k) E_n^-(\omega_k)$$
(4)

In our analysis, we considered a coherent transfer matrix defined as $T_{mn}(\omega_k) = \exp[i\varphi_{mn}(\omega_k)]/N$, where the phases $\varphi_{mn}(\omega_k)$ are random variables uncorrelated in space and Gaussian-correlated in frequency with a spec-tral correlation bandwidth $\Delta\omega_c$.[43,44] The spectral correlation is directly associated with the sample properties and inversely proportional to Thouless time (corresponding to the average confinement time of the field in the medium)[44]. In the presence of broadband illumination, the spectral correlation bandwidth $\Delta\nu_c$ is of critical importance, as it determines the total number of accessible spectral modes within the illumination bandwidth [9,36,45]. In our particular case, we convolve a white noise distribution with a Gaussian filter with a standard deviation of $\Delta\nu_c = \Delta\omega_c/2\pi = 400$ GHz along the frequency axis to impose a desired spectral correlation in the transfer matrix.

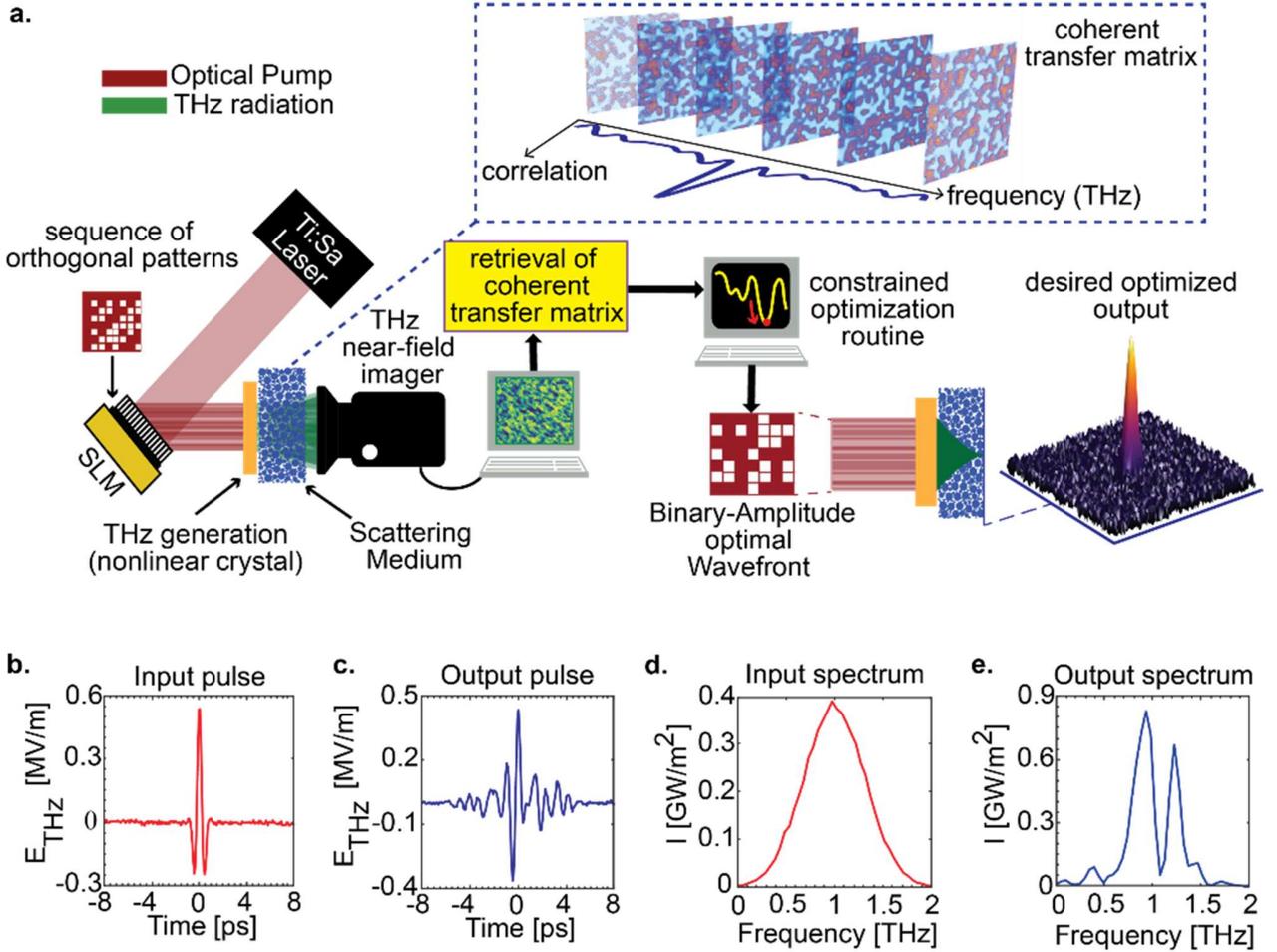

**Figure 1: Schematic of experimental driven methodology- a.** A conceptual overview of methodology, including the nonlinear conversion of optical patterns to THz structural waves and the retrieval of the transmission properties of the scattering medium in terms of a coherent transfer matrix. The full knowledge of coherent transfer matrix retrieved using an orthogonal set of patterns can be used to achieve scattering assisted focusing at the output of the scattering medium. **b.** Input THz pulse profile at the $n$-th pixel. **c.** Scattered THz pulse collected at the $m$-th pixel at the output end of the scattering medium. **d.** Intensity profile of input THz spectrum at $n$-th input pixel. **e.** Perturbation in output spectrum profile collected from the $m$-th pixel. For simulations, we considered 1 $nJ$ THz pulse of duration 250 $fs$ at the input with per pixel 40 dB SNR. The $3.2 \times 3.2\ mm^2$ sample illumination area is spatially sampled at 200 $\mu m$ resolution.

Figure 1a provides a conceptual overview of the THz-TDS experimental configuration we referenced in our modelling. An optical spatial light modulator (SLM) impressed a desired spatial pattern on an ultrafast optical field ($\lambda = 800nm$). This optical pattern is converted to a THz structured field via a nonlinear crystal, as discussed in Ref [38]. Without loss of generality, we assume a quadratic $\chi^{(2)}$ optical rectification process (e.g., ZnTe) that converts the optical intensity wavefront $\left|E^{optical}(x,t)\right|^2$ to a THz wavefront $E^{THz}(x,t)$ as follows:

$$E^{THz}(x,t) \propto \chi^{(2)}\left|E^{optical}(x,t)\right|^2 \qquad (5)$$

where $\chi^{(2)}$ is the second-order susceptibility of the nonlinear crystal. With this position, the THz field impinging on the scattering medium is defined in the frequency domain as $E^-(x',\omega) = \left|\tilde{E}^{optical}(x')\right|^2 f(\omega)$, where $f(\omega)$ is the spectrum of the THz pulse. The THz pattern impinges upon the scattering medium and produces a complex, time-dependent interference pattern at the output. Finally, a TDS image of the scattered THz wave is collected through a parallel, near-field imaging scheme based on electro-optical sampling [46]. To assess the robustness of our approach to experimental noise, we performed the theoretical analysis using a 40 dB Signal-to-Noise Ratio (SNR, per pixel) THz pulse that contains a white-noise term and water absorption signatures in the TDS trace and is compatible with experimental conditions. In Fig. 1b-e, we show an illustrative THz transmitted field as a function of the spatial and spectral coordinates obtained for plane wave illumination. The temporal profile of the pulse is significantly broadened (Fig. 1c), and the peak field is attenuated (Fig. 1e). When moving to the spectral domain (Fig. 1e), the transmitted THz field at a specific $m$-th pixel is characterized by a random modulation of its spectrum as a consequence of interference and dispersion effects induced by multiple scattering.

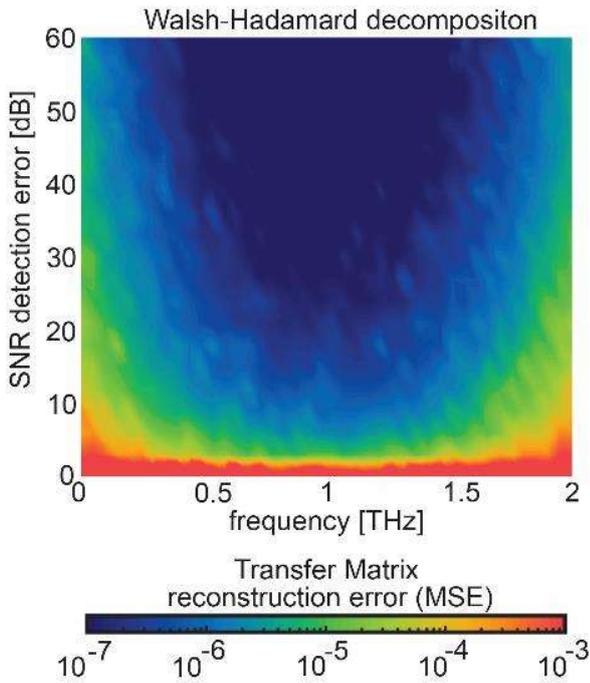

We reconstructed the coherent transfer matrix by acquiring the TDS output images corresponding to a pre-defined set of illumination patterns. In our simulations, we employed a Walsh-Hadamard decomposition scheme, i.e., we determined the full-wave responses corresponding to each column of an $N \times N$ Walsh-Hadamard matrix. We then extracted the frequency-dependent elements of the coherent transfer matrix through a linear inversion of the Walsh-Hadamard response (Fig. 2). The detailed reconstruction method is described in Supplementary Note 1.

**Figure 2: Transfer Matrix reconstruction**. Mean squared error (MSE) of the coherent transfer matrix elements as reconstructed through a Walsh-Hadamard decomposition.

The identification of an optimized optical spatial pattern $I_{opt}(x')$ that produces a given field profile of interest $Z_{target}(x_o, \omega)$ carries a few significant challenges. First, as can be easily evinced from Eq. (5), in optical rectification the THz field phase cannot be changed via optical phase changes, i.e., we can only control the amplitude of the THz patterns through the intensity distribution of the optical pump. Second, the spatial distribution of the optical intensity pattern is bound to be the same for all the different frequencies carried by the THz pulse. Due to these two constraints, we cannot invert the coherent transfer matrix directly, as the solution-pattern is likely a frequency-dependent amplitude and phase distribution. On the contrary, we must identify a single, amplitude-only field distribution that best approximates the desired field distribution at the output. It is essential to stress that this is a post-measurement process, as opposed to the case of typical optimization techniques relying on feedback loops between illumination and measurement. To this end, we cast our inversion problem in terms of a constraint least-square minimization of the following fitness function:

$$f\left[I_{opt}(x')\right] = \frac{1}{2}\left\|\int\int \widetilde{T_x}(x_o, x', \omega)I_{opt}(x')f(\omega)dx' - Z_{target}(x_o, \omega)\right\|_2^2$$

$$\text{subject to } 0 \leq I_{opt}(x') \leq 1, \ I_{opt}(x') \in \mathbb{R} \tag{6}$$

where $\|\cdots\|_2$ is the Euclidean norm. The constrained convex optimization problem defined in Eq. (6) can be solved using various techniques. We use the Trust-Region-Reflective algorithm, a well-established method capable of rapidly solving relatively large-scale problems with low memory requirements.[47] The ability to optimize the full-field properties of the transmitted field is a distinctive feature of this approach; Eq. (6) is, indeed, an absolute phase-sensitive optimization, corresponding to a field-driven best-fit rather than an intensity-driven fit.

## 4. Results and Discussion

### 4.1- Spatio-temporal focusing of THz wave through scattering medium

Our first objective is to invert the coherent transfer matrix to obtain a spatial-temporal localized focus spot at the output of the scattering medium, a classical scenario in the state-of-the-art. Such a task has been explored in the optical and infrared domain both for monochromatic[15,48] and ultrafast pulses[49,50], but never tackled for field sensitive systems and, in particular, for broadband THz fields. In our approach, the realization of a spatio-temporal focus corresponds to imposing the following target field profile in Eq. (6):

$$Z_{target}(x_o, \omega) = \delta(x_o - x_a)E_a f_a(\omega), \tag{7}$$

where $x_a$ is the desired focus position, and $E_a f_a(\omega)$ is the spectrum of the incident THz pulse. Equation (7) targets an output field localized in one spatial point with the same spectral profile as the incident pulse. The results are shown in Fig. 3a-c and effectively predict the formation of a sharp focus at the output. Quite remarkably, our amplitude-only optimized wavefront yields a spectral intensity enhancement ($\eta$) of approximately 16.76 (at 1 THz) at the focus spot (Fig. 3b). The field peak (Fig. 3c) is enhanced by a factor of 2.4, whereas the field-temporal standard deviation (the transient duration) is compressed by a factor of 3.5 with respect to the un-optimized case. It is worth stressing that, by observing Fig. 3c, not only the pulse is recompressed. As expected from a full-field function reconstruction, the field dynamics are reconstructed locally, similar to Fig. 1b. As discussed in Supplementary section 2, we verified that the optimal pattern is virtually identical to those obtained with standard iterative optimization techniques[51,52].

Since we are able to fully assess the transmission properties of the scatterer, our approach is easily extendable to more challenging conditions, including the formation of separate spatial foci with different spectral profiles. To this end, we generalized the target field profile from Eq. (6) to the case of two foci as follows:

$$Z_{target}(x_o, \omega) = \delta(x_o - x_a)E_a f_a(\omega) + \delta(x_o - x_b)E_b f_b(\omega) \tag{8}$$

where $x_a$ and $x_b$ correspond to the two different focus locations, and $E_a f_a(\omega)$ and $E_b f_b(\omega)$ denote two distinct spectral profiles, respectively. Our results are shown in Fig. 3d-3e. When considering identical spectral profiles at the output (Fig. 3d, $f_a(\omega) = f_b(\omega)$), we achieved the THz spectral intensity ~4 $GW/m^2$ ($\eta = 9.57$) and ~3.1 $GW/m^2$ ($\eta = 7.35$) at 1 THz, and a THz peak field enhancement of ~1.4 and ~1.6, respectively. When considering two different spectral profiles, centered at around 1.3 THz and 0.7 THz, respectively (Fig. 3e), the two foci exhibit a spectral intensity of ~2.5 $GW/m^2$ ($\eta = 5.60$) and ~1.8 $GW/m^2$ ($\eta = 3.96$), respectively.

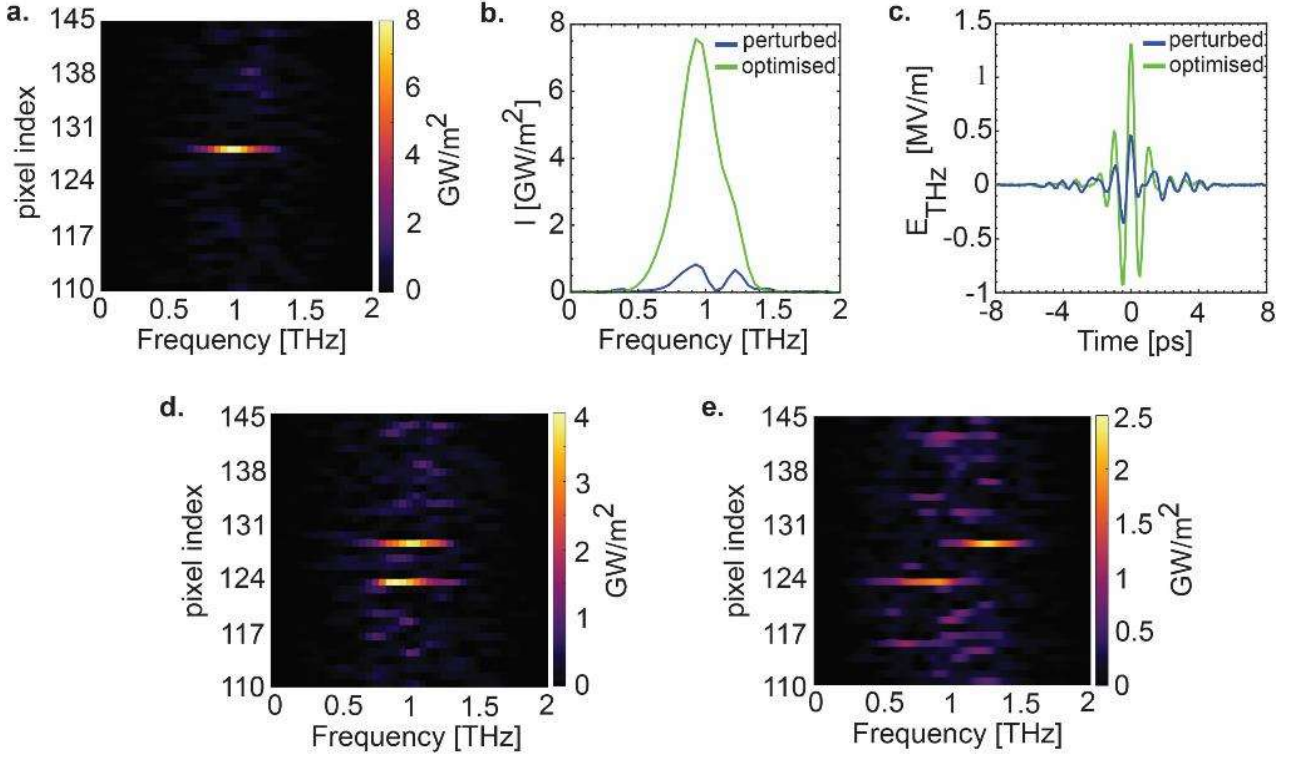

**Figure 3: Spatiotemporal focusing of THz field- a.** Optimized intensity distribution showing focus spot in THz band. **b.** Comparison between intensity profiles of perturbed THz spectrum (blue) and optimized spectrum (green) at the $m$-th pixel. **c.** THz pulse profile of scattered field (blue) and optimized field (green) from the $m$-th pixel of output plane. **d.** The intensity distribution of the output THz field showing two simultaneous focus spots at $m$-th pixel and $m'$-th pixel. **e.** The intensity distribution of optimized THz field for two simultaneous focus spots at $m$-th and $m'$-th pixel with two different spectrums centred around 0.7 THz and 1.3 THz. SNR per pixel: 40dB.

## 4.2 - Time-resolved retrieval of image object obscured by scattering medium

The reconstruction of the coherent transmission properties of the scatterer can also be directly applied to reconstruct the image of an object concealed by the scattering medium.[53-55] A particular possibility enabled by our methodology is the possibility of performing phase-sensitive reconstruction by leveraging our ability to reconstruct the full-wave properties of the transfer matrix elements. A numerical implementation of the image reconstruction process is shown in Fig. 4, where we place a phase mask $U(x')$ between the generating crystal and the scattering medium. In the frequency domain, the corresponding transmitted field reads as follows:

$$M(x_o, \omega) = \int \widetilde{T_x}(x_o, x', \omega) \, exp[iU(x')]E^-(x', \omega)dx' \qquad (9)$$

where $M(x_o, \omega)$ is time-Fourier transform of the space-time measurements. To retrieve the original image from the measurements, we perform a standard deconvolution of the transfer matrix that yields the time-resolved image ($I_{retrieved}$) as,

$$I_{retrieved}(x,t) = \mathcal{F}^{-1}\left\{\left[\widetilde{T_x}(x_o, x', \omega)\right]^{-1} * M(x_o, \omega)\right\} \tag{10}$$

where $(*)$ denotes a spatial convolution, $\mathcal{F}^{-1}$ is the inverse time-Fourier transform, and $[...]^{-1}$ is the inversion operator. As is customary in deconvolution problems, the main task lies in finding out the inverse of $\widetilde{T_x}(x_o, x', \omega)$. We applied the Moore-Penrose pseudo inversion method, implemented through a truncated singular-value-decomposition[56]. As shown in Fig. 4c, before applying the deconvolution routine, the THz pulses corresponding to the two distinct pixels are thoroughly perturbed, representing the multiplexing of waves due to multiple scattering. Fig. 4d shows the waves corresponding to two separate pixels (red and cyan dot) after the deconvolution process. We calculated the Structural Similarity (SSIM) index[57] to quantify the quality of the reconstruction process, as shown in Fig. 4e. SSIM values obtained at -0.52 ps, 0 ps and 0.12 ps in our time reference are 0.20, 0.83

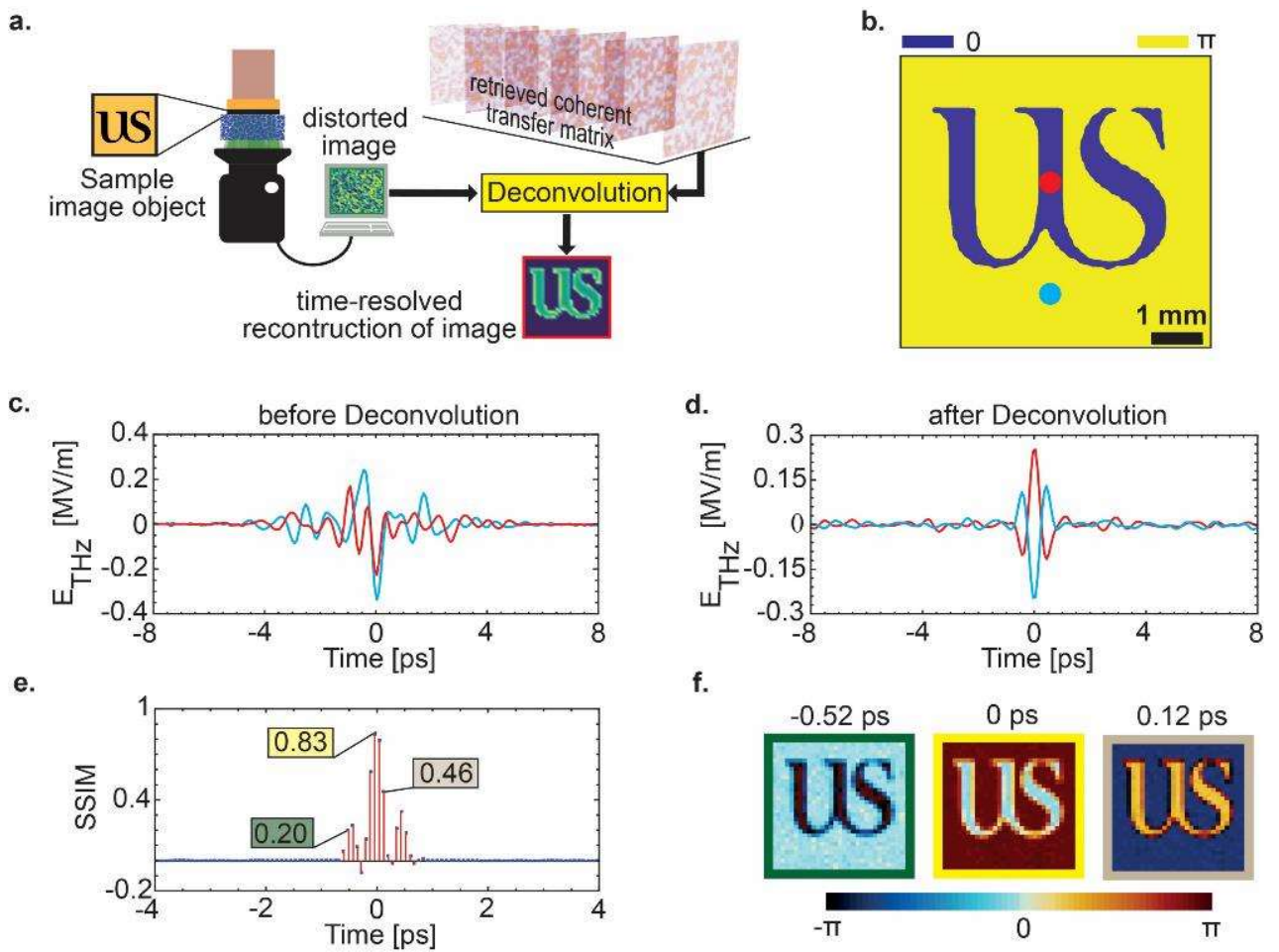

**Figure 4: Time-resolved THz phase-sensitive imaging through the scattering medium. a.** Schematic of imaging methodology **b.** The phase image object. **c.** Temporal evolution of output speckles corresponding to two different pixels (red and cyan dot shown in **b**) before deconvolution. **d.** Temporal evolution of reconstructed THz pulse (after deconvolution) for two different pixels (red and cyan dot). **e.** Structural Similarity (SSIM) index in the time-resolved reconstruction of an image object. **f.** Reconstructed phase images at t= -0.52 ps, 0 ps, 0.12 ps. The $6.4 \times 6.4\ mm^2$ sample illumination area is spatially sampled at 200 $\mu m$ resolution. Logo used with permission from the University of Sussex. (see time-resolved imaging: video-1)

and 0.46, respectively, showing high fidelity in the time-resolved reconstruction of the image. The specific reconstruction results shown in Fig. 4f are the fixed-time reconstructed phase images of the transmitted field at the same time values. Analogous results for an amplitude-only object (i.e., a metallic mask) are included in Supplementary Fig. 3. As a final example, we extended our image reconstruction approach for the THz vortex beam[58] and simulated the spatio-temporal field-phase profiles for the $L_0^1$ and $L_1^1$ THz vortex beam (see Supplementary Figure 4). We deconvolved the spatial field-phase information of $L_0^1$ and $L_1^1$ THz vortex beam with the complex scrambled output obtained from propagation through the scattering medium and shown their retrieved spatial field-phase profiles at t=0 ps (Fig. 5a, 5b and Fig. 5c, 5d). Figures 5e and 5f are the temporal profile of the retrieved THz field corresponding to two different pixels (green and cyan dot).

## 5. Conclusions

In this work, we have theoretically demonstrated a deterministic approach towards the coherent spatiotemporal control of THz waves propagating through a scattering medium. Our methodology combines the nonlinear conversion of optical patterns to THz structured fields with field-sensitive THz field detection, as enabled by state-of-the-art TDS technology. We have shown how the full-wave detection of the scattered THz field enables retrieving the field-sensitive transfer function of the medium directly in a deterministic fashion, as described through a coherent transfer matrix modelling. We sample the complex time-domain elements of the coherent transfer matrix by projecting a sequence of orthogonal Walsh-Hadamard patterns. The TDS allows for a sufficient description of the coherent transfer matrix to ena-ble spatio-temporal control through a direct inversion approach. We identified the spatial profiles that yield a desired output field distribution through a convex constraint optimization routine compatible with real-life experimental conditions. As relevant examples, we demonstrated the formation of single and multiple spatiotemporal foci and the retrieval of complex field distributions and phase-only images concealed by the scatterer. Our results suggest that it is still possible to investigate scattering unaffected open-path via a time-domain deterministic approach in an experimental-driven constrained scenario. Such a control could have a profound impact, especially for THz imaging, where wave-shaping is generally a challenge. In addition, we envision a role in Time-resolved characterization techniques of complex media, including deep-tissue biological imaging.

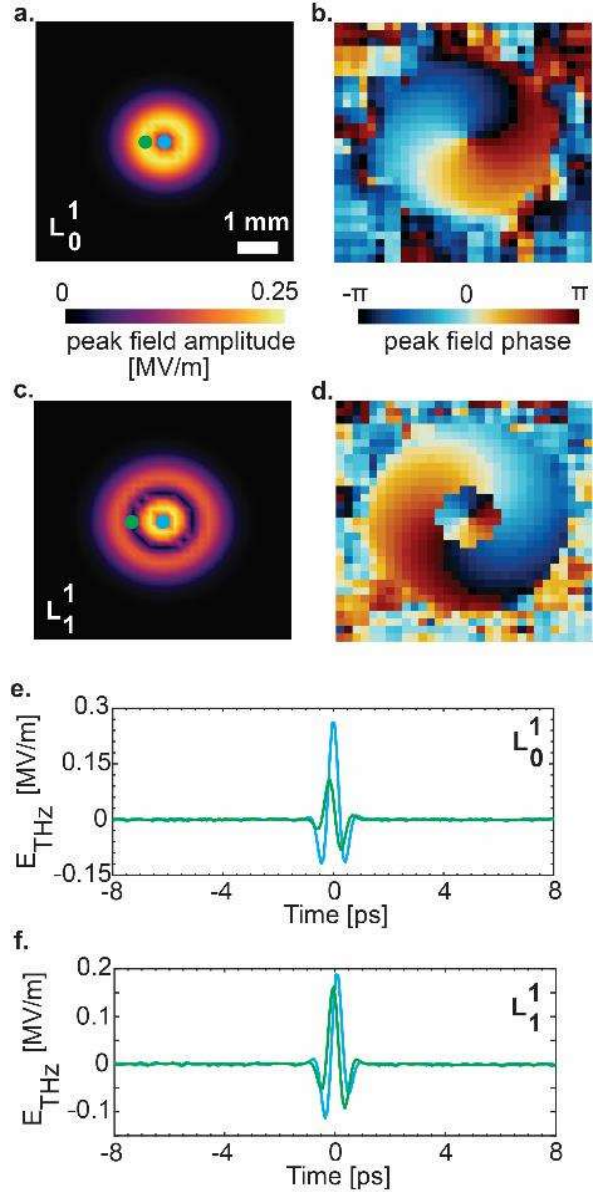

Figure 5: Complex propagation of THz vortex beam through the scattering medium. **a-b.** Retrieved spatial field and phase distribution of $L_0^1$ vortex beam at t=0 ps. **c-d.** Retrieved spatial field and phase distribution of $L_1^1$ vortex beam at t=0 ps. **e-f.** Temporal profile of retrieved THz pulse corresponding to two different pixels (cyan and green dot). The $6.4 \times 6.4\ mm^2$ sample illumination area is spatially sampled at 200 $\mu m$ resolution. (see video 2).

ASSOCIATED CONTENT

The Supplemental Materials, including Reconstruction videos for image and THz vortex beam are available.

AUTHOR INFORMATION

Corresponding Author

* E-mail: m.peccianti@sussex.ac.uk

**Present Addresses**

* Emergent Photonics Lab (EP*ic*), Department of Physics and Astronomy, University of Sussex, Brighton, BN1 9QH, U.K.

**Author Contributions**

All the authors engaged in the general discussion regarding the basic science of the paper. V.K performed the calculation reported. All the authors contributed to the general understanding of the results and to the drafting of the paper. J.S.T.G. and M.P supervised the general research activities.

**Funding Sources**

This project received funding from the European Research Council (ERC) under the European Union's Horizon 2020 Research and Innovation Programme Grant No. 725046. We acknowledge financial support from the (UK) Engineering and Physical Sciences Research Council (EPSRC), Grant No. EP/S001018/1 and EP/T00097X/1 and the Leverhulme Early Career Fellowship ECF-2020-537.

**Notes**

The authors declare no competing financial interest. The data sets for all figures are freely accessible at DOI.

ACKNOWLEDGMENT

V.K, V.C. and L.P. acknowledge the support from the European Research Council (ERC) under the European Union's Horizon 2020 Research and Innovation Programme Grant No. 725046. J.S.T.G acknowledges the support the Leverhulme Early Career Fellowship ECF-2020-537.

ABBREVIATIONS

THz, terahertz; TDS, time-domain spectroscopy; FROG, Frequency resolved optical gating; SPIDER, spectral phase interferometry for direct electric-field reconstruction; SLM, spatial light modulator; ZnTe, Zinc telluride; SNR, signal to noise ratio; SSIM, structural similarity index.

*Supplementary Information*

# Deterministic THz wave control in scattering media

Vivek Kumar[1], Vittorio Cecconi[1], Luke Peters[1], Jacopo Bertolotti[2], Alessia Pasquazi[1], Juan Sebastian Totero Gongora[1], Marco Pecanti[1,*]

[1] Emergent Photonics Lab (EPic), Department of Physics and Astronomy, University of Sussex, Brighton, BN1 9QH, U.K.

[2] Department of Physics and Astronomy, University of Exeter, Exeter, Devon EX4 4QL, UK.

[*m.pecanti@sussex.ac.uk](mailto:m.pecanti@sussex.ac.uk)

### Supplementary Note 1. Reconstruction of the coherent transfer matrix

The reconstruction of the coherent transfer matrix can be generally performed by sampling the input-output relation with a series of orthogonal functions. The Walsh-Hadamard basis represents a canonical example known to provide a higher signal to noise (SNR) ratio compared to single-pixel, raster-scan illumination.[1] The general form of the coherent transfer matrix at particular frequency $v$ is given by:

$$T_{mn}(v) = \begin{bmatrix} t_{11(v)} & \cdots & t_{1n}(v) \\ \vdots & \ddots & \vdots \\ t_{m1}(v) & \cdots & t_{mn}(v) \end{bmatrix}, \tag{S1}$$

where, $t_{ij}(v)$ is the complex-valued field propagator that connects $j$-th input mode and $i$-th output pixel. In the presence of broadband illumination, the incident field for a given pattern is expressed as

$$E_n^-(v) = H_p f(v), \tag{S2}$$

the $p$-th column vector

$$h_p^{\pm} = \frac{1}{2} \pm \frac{1}{2} H_p. \tag{S3}$$

With this approach, the corresponding output to each binary pattern $c_p^+(v)$ and $c_p^-(v)$ can be acquired by performing the Fourier transform of the time-resolved measurements obtained from TDS and, in an analogous fashion to differential ghost-imaging approaches, the coefficient corresponding to the $p$-th Walsh-Hadamard pattern is simply expressed as:

$$c_p(v) = c_p^+(v) - c_p^-(v) \tag{S4}$$

For each frequency, the differential signals are stacked into a measurement matrix $M(x,v) \in \mathbb{C}^{N \times N}$, and the transfer matrix elements can be obtained by a linear inversion of $M(x,v)$ for each frequency. In Figure 2 of the main text we quantitatively estimated the robustness of our approach against additive noise at the detection by computing the Mean Square Error (MSE) of our reconstruction for different values of the detection SNR. The MSE is defined as:

$$MSE(v) = \frac{1}{N} \| T_{reconstructed}(v) - T_{mn}(v) \|_F^2 = \frac{1}{N} \sum_m \sum_n \left| T_{reconstructed,mn}(v) - T_{mn}(v) \right|^2 \tag{S4}$$

where $\| ... \|_F$ is the Frobenius norm. We compared the MSE values for raster scan and Walsh-Hadamard decomposition (Fig 2 main text). As a direct comparison, in Supplementary Figure 1 we report the same analysis for a single-pixel, raster scan measurement of the transfer matrix elements for different values of the detection SNR.

Supplementary Figures

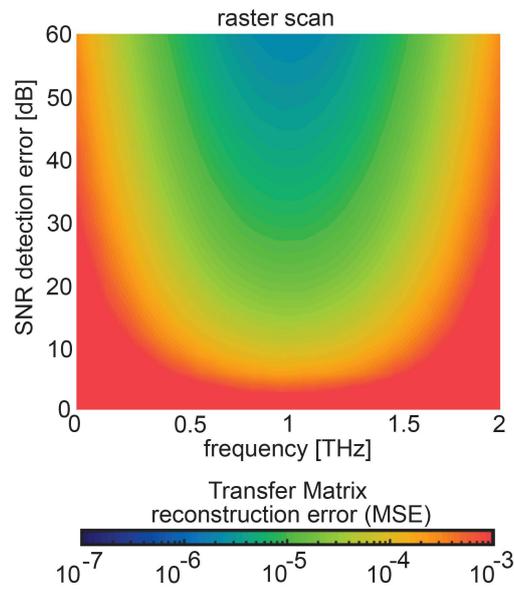

**Supplementary Figure 1:** Mean Square Error in retrieval of coherent transfer matrix using raster scan.

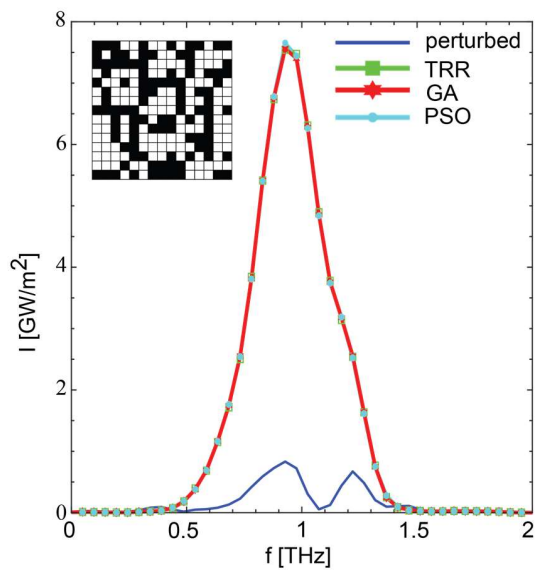

**Supplementary Figure 2.** Performance analysis of spatio-temporal control of THz wave using Trust-Region-Reflective (TRR), Genetic algorithm (GA) and Particle Swarm Optimisation (PSO) for constraint inversion of coherent transfer matrix. Optimized spectral intensity at the focus spot obtained from TRR, GA and PSO (blue: scattered intensity, green square: using TRR, red star: using GA and cyan dot: using PSO). TRR, GA, PSO are conversing to the same binary-based pattern solution as shown in the Inset: Optimized binary-based pattern (black: 0, white: 1).

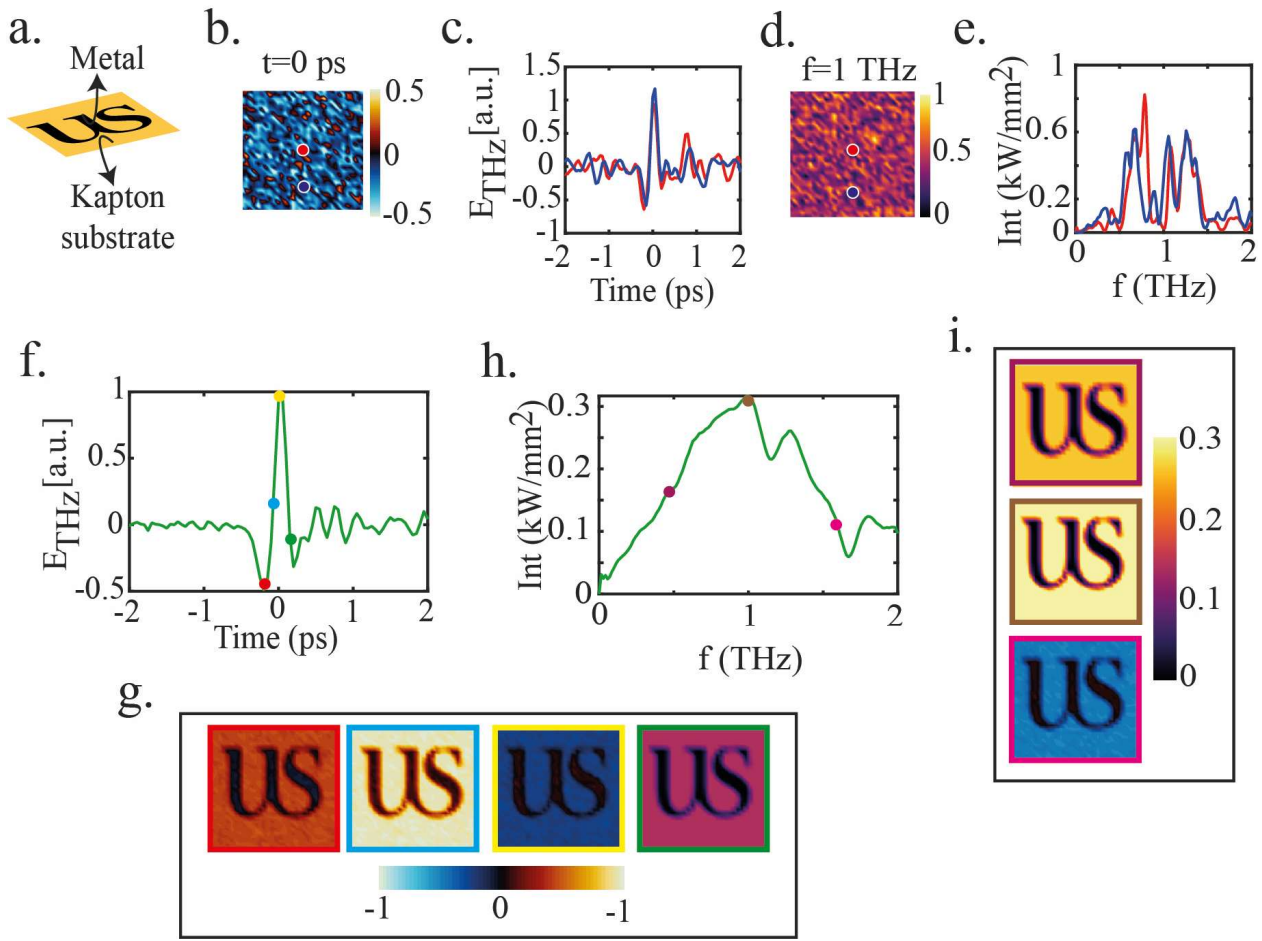

**Supplementary Figure 3: Hyperspectral THz imaging through the scattering medium**. **a.** Schematic of metallic Image object. **b.** Interference pattern formed at the output after propagation through the scattering medium. **c.** Temporal evolution of output speckle corresponding to two different pixels. **d.** Spectral intensity distribution of scrambled image object propagated through medium at 1 THz. **e.** Intensity profile of output speckle corresponding to two different pixels. **f.** Temporal evolution of reconstructed image object averaged over all the pixels. **g.** Fix time retreival of images at -0.31 ps, -0.26 ps, 0 ps and 0.17 ps. **h.** The spatially averaged reconstruction of THz spectrum. **i.** Retirved hyperspectral images at 0.55 THz, 1 THz and 1.55 THz. The $6.4 \times 6.4$ $mm^2$ smaple illumination area is spatially sampled at $200$ $\mu m$ resolution. Logo used with permission from the University of Sussex.

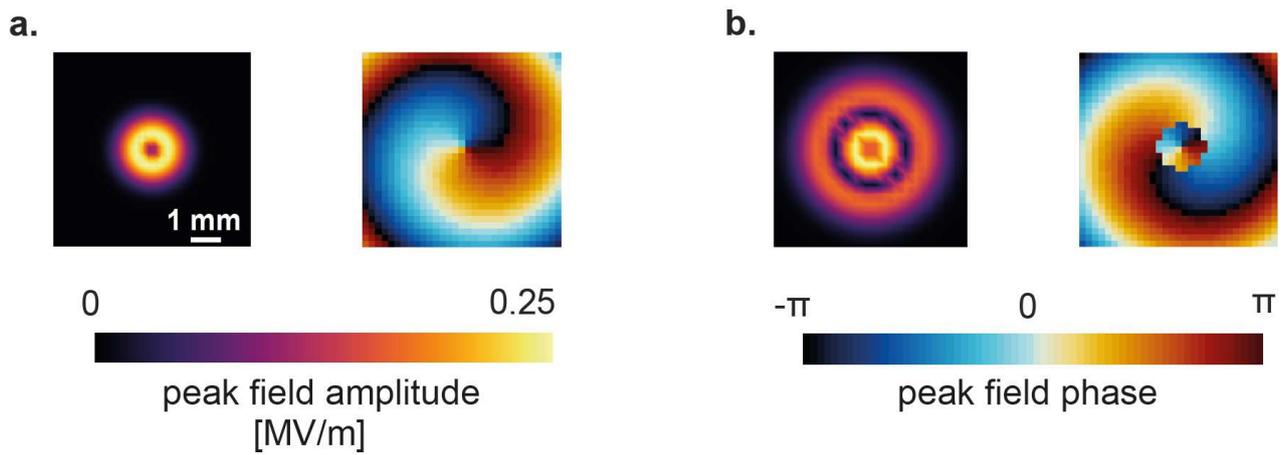

**Supplementary Figure 4: Simulated $L_0^1$ and $L_1^1$ THz vortex beams. a.** Spatial field and phase distribution for radial index 0 and topological charge 1. **b.** Spatial field and phase profiles for radial index 1 and topological charge 1. THz vortex beams are plotted at the peak of the THz pulse after propagation of 0.5 mm in free space. The $6.4 \times 6.4 \ mm^2$ spatial area is sampled at